\documentstyle[12pt,aaspp]{article}

\doublespace

\input{psfig}

\def\grad{{\bf\nabla}}

\def\hb{{\bar{h}}}

\def\ub{{\bar{u}}}

\def\del{{\partial}}

\begin{document}

\title{Buoyancy driven rotating boundary currents}

\author{P.A. Yecko \& S.P. Meacham}

\affil{Physics Department, University of Florida}
\affil{and}
\affil{Department of Oceanography, Florida State University}

\vskip 1cm
\section{Introduction}

\noindent
Buoyancy is responsible for the formation of many important
geophysical and astrophysical flows.   In a rotating system, a
sustained source of positive or negative buoyancy may give rise to an
extended horizontal current against a lateral boundary, a gravity
current.  In such a current, the Coriolis force associated with the
horizontal flow in the current is balanced by an offshore pressure
gradient supported by the wall.  
The resulting flow is then
bounded by a solid boundary on the right, (looking along 
the direction of flow in the Northern hemisphere) 
and by a front on the left.  If the wall is sloping rather than
purely vertical, the component of the Coriolis force parallel to the
wall in the vertical plane can balance the similar component of the
gravitational force acting on the current.
Where the source is persistent, such currents can reach states of
quasi-equilibrium.  See Griffiths~$^1$ for a review
of boundary gravity currents.

It has been noted that the instability of many varieties
of sheared flows cannot be determined locally~$^2$.  
As an illustrative example, recall that for a two-dimensional
incompressible shear flow, $U = U(y)$, 
one {\sl must} use boundary conditions appropriate to the
corresponding eigenvalue problem in order to obtain 
a necessary condition for instability.  That condition is
simply the vanishing of the vorticity gradient ($U_{yy} = 0 $)
somewhere in the flow -- the celebrated Rayleigh criterion.
Local approximation leads to a different condition and
the erroneous prediction of likely instability~$^3$.

Investigation of more complicated (compressible) 2-D shear 
flows has confirmed the nonlocality of wave-interaction, or
resonant, instabilities~$^4$ and more specific studies of
boundary current configurations which account for the
coupling of the current to its environment have also 
found that mode resonances can alter the intrinsic stability
characteristics of flows~$^{5,6,7}$.

\section{The Meddy problem}

\noindent
How can the presence of warm, salty Mediterranean water be explained when
it is found in the Western Atlantic ocean?  The Mediterranean water is
delivered in compact lens-shaped vortices affectionately known as
{\it Meddies}.  
Such long-range mixing capabilities of some currents
has a profound influence on the global ocean circulation.  This 
example poses two distinct questions:

\vskip .2cm
 {\bf 1}  {\sl How are Meddies formed?}

 {\bf 2}  {\sl How do they subsequently propagate ?}

\vskip .2cm


The propagation of coherent patches of vorticity is only partially
understood.  A number of different long-range mechanisms can affect the
motion of vortices.  These include: interactions between different
vortices, advection by a large scale background flow, interaction with
a background vorticity or potential vorticity gradient (a good example
of this is the ``planetary beta'' effect seen in flows in a thin shell
of fluid on the surface of a rotating spherical planet~$^8$),
interaction of a vortex with a horizontal or vertical boundary, small
scale dissipation within and around a vortex, and MHD effects in
examples of astrophysical vortices.

\section{Laboratory experiments}

\noindent
To investigate question 1, we developed a series of laboratory
experiments to study buoyancy-driven currents in rotating stratified
environments.  Although we have targeted a specific oceanographic 
flow to model, it is tantalizing to consider that a reduced
problem such as the one we propose here may also be relevant to
other problems.  In binary star systems, to consider one
possible example, infall
material from one star can feed an accretion disk around the
other; the subsequent equilibration of the new material may well
lead to buoyancy driven currents of a similar nature.
But this study is primarily a model for the Mediterranean outflow,
where we have the benefit of comparison with direct observations.
The two principal goals of the experiment were:

\vskip .2cm
(a) to determine the structure (velocity shear and thickness) of the
currents initially formed from a negatively buoyant inflow

(b) to examine the stability properties of these current

%

\vskip .2cm

A diagram of he experimental apparatus is shown in figure 1.  The
interface of the stable ambient two-layer stratification intersects
the surface of a conical section at mid-depth.  A thymol blue solution
of intermediate density is held outside the tank and also fills a
reservoir at the tank's outer edge.  The densities of the overflow
water ($\rho_2$) and the bottom layer ($\rho_3$) were achieved by
dissolving sugar in distilled water.

A current is initiated by establishing a flow that causes overflow in
the reservoir and the subsequent buoyant descent of the overflow water
down the sloping surface toward the layer interface, where it becomes
neutrally buoyant.  Because of the rotation, the descending current
turns to flow cyclonically along the topographic slope. Except for
extreme initial conditions, the current will still reach the layer
interface.  Along one radius of the conical surface, an array of fine
(0.006'' diameter) vertical wires are placed at 0.44 cm intervals.
The wires are pulsed with an electrical current to activate thymol blue
tracer in fluid of the current, allowing direct visualization.  
A camera is positioned above the tank
(where the length of advected tracer is used to determine velocities)
and a mirror is oriented to allow a head-on view of the current (where
the height of unadvected tracer determines local thickness).  Crude
evaluation of the response in the ambient layers is done by tracking
floating and, where possible, submerged tracers.  Images captured with
the CCD camera were digitally stored using a video framer and the
measurement of tracer was facilitated by enhancing contrast using
standard algorithms.  Pre-marked intervals on the experimental
surfaces provided accurate distance standards, and the real-time
record kept on video simplified measurement of time intervals.

\centerline{\psfig{figure=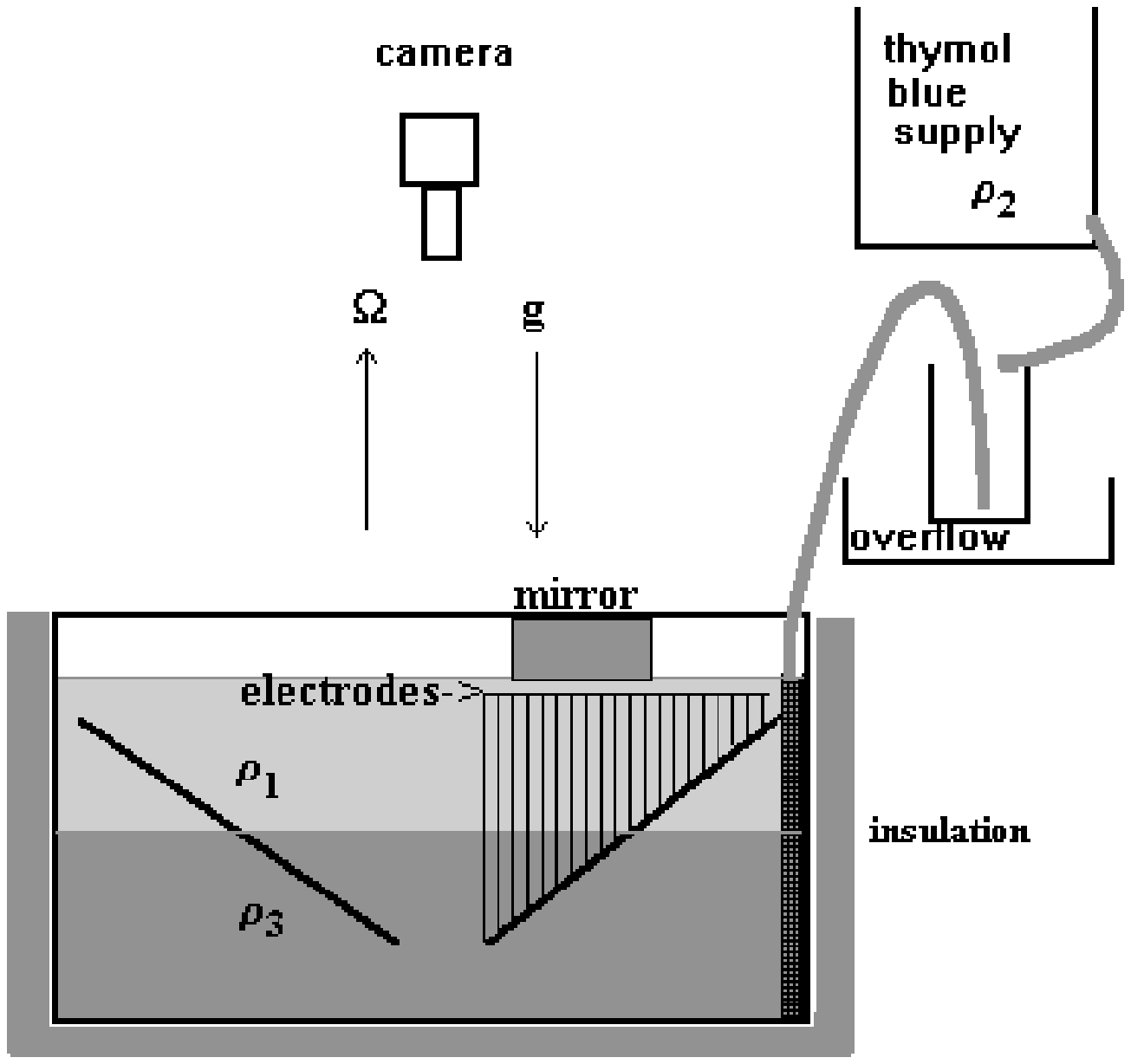,height=5.in}}

\noindent
{Figure 1. The experimental apparatus: the cylindrical tank and 
 the submerged conic slope are shown in cross-section; the tank has
 diameter of $\sim 60$ cm.  Everything sits on a rotating table.
 }


In all experiments, the densities were prepared with the values:
$\rho_1 = 1.0$g/cc, $\rho_2 = 1.005$g/cc, and $\rho_3 = 1.01$g/cc;
the depths of the ambient layers were $11$cm each. 
The reduced gravity between layers then took the values: $g'_{12} =
g'_{23} = 5.0$g cm/sec$^2$; while $\Omega$ was either $1.$ or $1.5$.   
The principal control value was the volume flux, $\cal{S}$, of
$\rho_2$-fluid, which was varied from $0.15 - 1.0$cc/sec.  $\cal{S}$
is directly related to a characteristic current depth $H$, which
varied from $1 - 3$cm.  Two values of the topographic slope, $\alpha$,
were used, 0.0 (a vertical wall) and 0.25. 

From the parameters above, we can construct several useful quantities. 
A Rossby deformation length is given by $L_D = (g'H)^{1\over
2}/2\Omega$, and the ratio of the current width to $L_D$ is indicative
of the relative importance of baroclinic and barotropic disturbances.
%
%
A Reynolds number is given by $Re=(g' H^3 / \nu^2)^{1\over 2}$, and for
these experiments reached nearly $10^3$, but
a more useful quantity is the Froude number $F=U^2/(g'H)$.

These and other  experiments 
have verified that baroclinic currents and instabilities of interest are characterized not so much by $Re$ as by $F$.
In oceanic flows, $F$ can vary between $10^{-2}$ and $10^2$, and in the
laboratory we can easily produce a range of $F$ between $10^{-2}$ and
$10$, simply by forming velocities $U$ in the range $0.1$ to $1$ cm/sec.

In high Reynolds number flows without ambient stratification,
experiments have shown that instabilities of Kelvin-Helmholtz type
typically grow rapidly at the nose, while further upstream -- where
the steady current is well established -- baroclinic instabilities
produce eddies.  We focus on the established current rather than its
nose and the details of the intrusion process.  For this reason, we
initiate the current with a small density enhancement to minimize
the initial Kelvin-Helmholtz mixing, while using a tank of
sufficient horizontal dimension to traverse the relevant Froude number
range.

\section{Steady three layer currents}

What is the three-layer, steady, streamwise uniform solution?
An arbitrary three layer arrangement of fluid is described by the
familiar shallow layer system:
\begin{equation}
{d{\bf u}_1\over dt} + {\bf f}\times{\bf u}_1 = 
 -g\grad (h_1 + h_2 + h_3) \; ,
\end{equation}
\begin{equation}
{d{\bf u}_2\over dt} + {\bf f}\times{\bf u}_2 = 
 -g{\rho_1\over\rho_2}\grad h_1 - g \grad (h_2 + h_3) \; ,
\end{equation}
\begin{equation}
{d{\bf u}_3\over dt} + {\bf f}\times{\bf u}_3 = 
 -g{\rho_1\over\rho_3}\grad h_1 - g{\rho_2\over\rho_3}\grad h_2
 -g\grad h_3 \; ,
\end{equation}
and
\begin{equation}
{\del h_i\over\del t} + \grad\cdot\left(h_i{\bf u}_i\right) = 0 \; ,
\end{equation}
holds for each layer $i=1,2,3$.

We will let $Ox$ be the streamwise direction, parallel to the boundary,
and $Oy$, the cross-stream direction.
For a boundary current of intermediate density ($i=2$) flowing against
a vertical wall, a steady, streamwise uniform current with velocity
$\ub(y)$ and thickness $\hb(y)$ must satisfy
potential vorticity and momentum conservation.:
\begin{equation}
{d\over dt}\left({f - \del_y\ub \over \hb}\right) = 0 
\;\;\;\;  \& \;\;\;\; 
f\ub = - g^* \del_y \hb
\end{equation}
where the reduced gravity here is $ g^* = g {(\rho_2-\rho_1)(\rho_3 
- \rho_2)\over \rho_2 (\rho_3 - \rho_1)} $.

\vskip .5cm
\noindent
For uniform {\it non-zero} initial potential vorticity, $Q_2(t=0)=Q_*$,
we find currents of width $L$ having:
\begin{equation}
\ub(y)=fl{\sinh(y/l)\over \cosh(L/l)} \;\;\;\;\; \& \;\;\;\;\;
\hb(y)=h_*-h_*{\cosh(y/l)\over \cosh(L/l)} \; ,
\end{equation}
where a deformation length is now associated with the initial
conditions:
\begin{equation}
l=\left({g^* \over fQ_*}\right)^{1/2} \; .
\end{equation} 

To extend the solutions above to the laboratory configuration, in 
which the boundary along which the current flows has some slope,
we must simply match the solution above to a two layer solution of
a current flowing along a slope (this sub-problem is the subject
of several previous studies, see the review~$^1$ for references).

\section{Results}

\noindent
Steady solutions are shown in figure 2 alongside measured profiles.
Experiments performed with vertical boundaries have found the strong
tendency of currents to be unstable, even when the currents were
initiated from a uniform potential vorticity source.  This instability
set in uniformly along the current once it became sufficiently
wide, forming a chain of eddies.  The eddies were roughly the same
width as the current and moved irregularly following separation.

\centerline{\psfig{figure=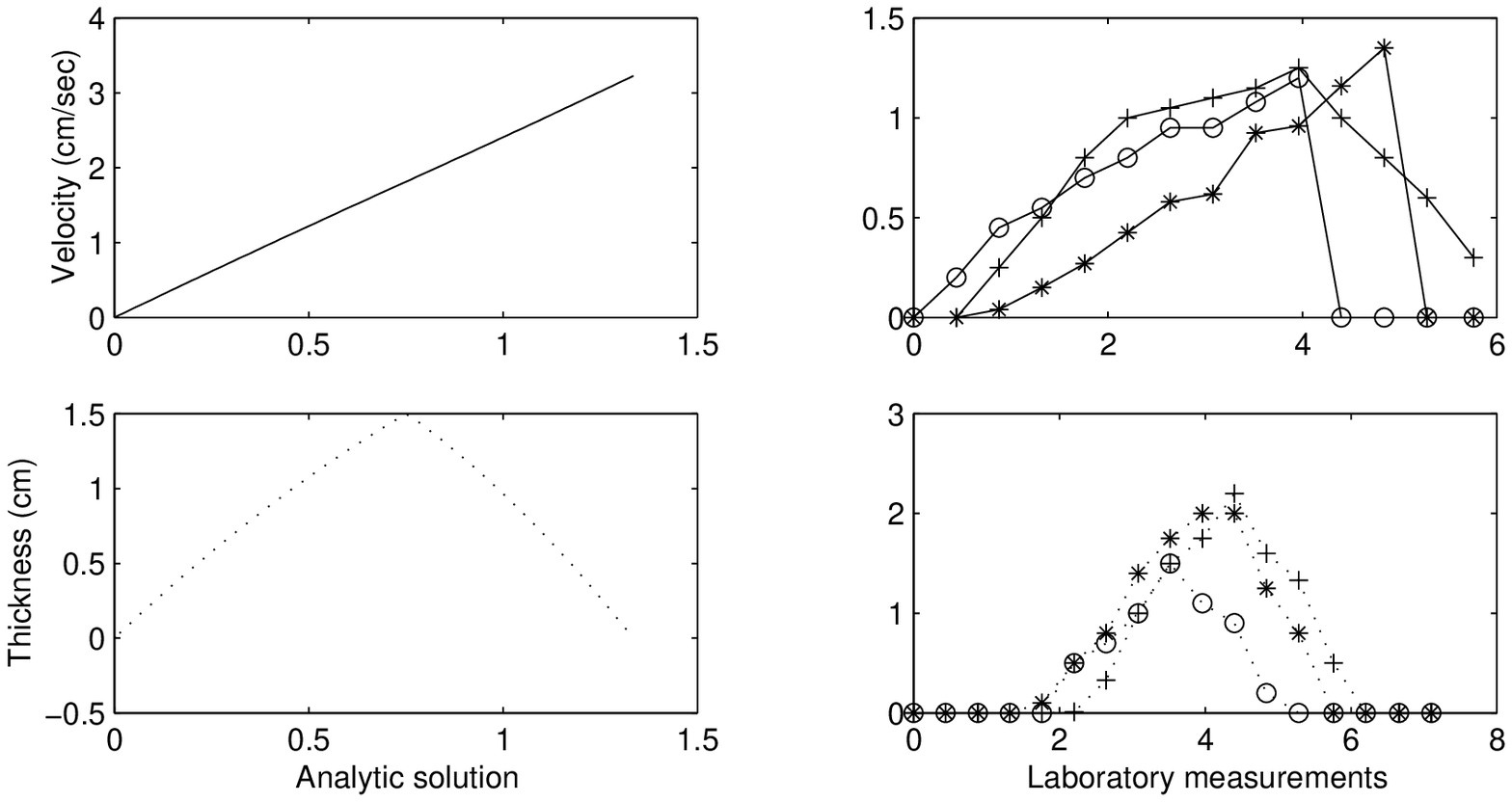,height=3.5in}}

\noindent
{Figure 2. (left) Steady solutions for u-velocity and thickness of the boundary
current in the three-layer system flowing along a sloping boundary; (right) The
measured u-velocities and thickness for three experiments, one case conincides
with the shedding of a cyclone.
 }

Analogous experiments along sloping boundaries have found the 
currents to be stable over much longer times and for wider 
currents.  The stabilization of such currents by the presence of
a slope should be emphasized -- it suggests that the instabilities
found for highly idealized and often symmetrical posed problems
may not be realized in more general settings.  The first step in
proving or disproving this hypothesis is understanding
the steady current configurations and a subsequent stability
analysis.  What we have reported here are first measurements of
steady current profiles for these stable cases.

Related experiments, in which a surface current is allowed to
eventually encounter a sloping boundary, have also found that a slope 
can have a stablizing effect -- following the
encounter, the current was found to be slower, wider, and more laminar
than it would be along a vertical wall$ ^9$.

A final point concerns the formation of Meddies:
in the process of initiating experimental currents,
the intermediate density fluid formed two distinct
equilibria.  The first as it flowed through less dense water,
down the slope and to the right; after reaching the 
layer interface with the densest water, it spread
to the right and along the sloping boundary.
Both stages were rather laminar and steady, but at
the transition, a single large cyclone was typically
shed.  The Mediterranean undergoes a similar transition
where it negotiates a sharp turn at Cape St.~Vincent 
and presumably
other significant topographic variations,
such as channels, are also encountered.  
The role of such localized perturbations
may help explain why, when experimental currents suggest
stability, their oceanic counterparts are able to form
eddies.

The excitation of inertial waves in the ambient
homogeneous fluid is also found.

Finally, we would like to use these preliminary results to
initiate a more thorough investigation of the linear instability 
in the full three layer problem as well as the basis for a 
companion set of numerical experiments to duplicate those done
in the laboratory.

\vfill
\eject
\vskip 1cm
\centerline{\bf Acknowledgements}

The authors would like to thank the National Science Foundation for
supporting this project as part of contract OCE-9301318, and
Dr. Ruby Krishnamurti for being of invaluable assistance in the lab;
and finally, the support and hospitality of the 
Geophysical Fluid Dynamics Institute at FSU.


\vfill
\eject

\vskip 2cm

\vfill
\eject

\vskip 2cm

\end{document}